\documentstyle[12pt]{article}
\def\prd#1{{\em Phys.~Rev.}~{\bf D#1}\ }
\def\prl#1{{\em Phys.~Rev.~Lett.}~{\bf #1}\ }
\def\plett#1{{\em Phys.~Lett.}~{\bf #1B}\ }
\def\np#1{{\em Nucl.~Phys.}~{\bf B#1}\ }
\def\deg{\ifmmode{^{\circ}}\else ${^{\circ}}$\fi}
\def\ni#1{\noindent$(#1)\quad$}

\def\gsim{\,\raisebox{-0.13cm}{$\stackrel{\textstyle>}{\textstyle\sim}$}\,}
\def\lsim{\,\raisebox{-0.13cm}{$\stackrel{\textstyle<}{\textstyle\sim}$}\,}
\def\bi{\begin{itemize}}
\def\ei{\end{itemize}}
\def\ed{\end{document}}
\def\be{\begin{equation}}
\def\ee{\end{equation}}
\def\beq{\begin{eqnarray}}
\def\eeq{\end{eqnarray}}

\def\lam{\ifmmode{\Lambda}\else $\Lambda$\fi}
\def\s{\ifmmode{S}\else $S$\fi}
\def\t{\ifmmode{T}\else $T$\fi}

\def\mod{{\cal M}}
\def\Us{U^*}
\def\Ys{Y^*}
\def\YY{\Ys Y}
\def\uu{\Us U}
\def\tfrac#1#2{{\textstyle\frac{#1}{#2}}}

\def\gev{\ \mbox{GeV}}

\def\pri{^{\, \prime}}

\def\eps{\epsilon}
\def\dvt{\Delta V_T}
\def\pij{\Pi^i_j}
\def\ms{m_{3/2}}

\def\eb{\end{thebibliography}}

\def\eqn#1{Eq.~(\ref{eq:#1})}
\begin{document}
\begin{titlepage}
\begin{flushright}  {\sl NUB-3120/95-Th}\\
{\sl May 1995}\\
hepph/9506056
\end{flushright}
\vskip 0.5in
\begin{center}
{\Large\bf Can Hidden Gauginos Form Condensates?}\\[.5in]
{Haim Goldberg}\\[.1in]
{\sl Department of Physics}\\
{\sl Northeastern University}\\
{\sl Boston, MA 02115}
\end{center}
\vskip 0.4in
\begin{abstract}
Supersymmetry breaking may be linked to the formation of
gaugino condensates in a hidden  sector.
In this work, the process of formation of the condensate is examined in a
cosmological context, using an effective field theory of the gaugino
bilinear which provides a reasonable interpolation between the high- and
low-temperature phases.
The implementation of anomaly requirements generates a large potential barrier
between the zero-condensate configuration and that of the true (SUSY-breaking)
vacuum. As a consequence, the transition to bubbles of true vacuum
may be  subject to an enormous
exponential suppression. This leads to the same difficulties with inhomogeneity
of the universe which occurred in the original inflationary scenarios.
\end{abstract}
\end{titlepage}
\setcounter{page}{2}
\section{Introduction}

Softly-broken supersymmetry constitutes a reference point of much of
present day
discussion of elementary particle physics. Yet a well-defined dynamical
mechanism driving the soft-breaking remains an elusive goal for theorists. An
extensively studied possibility is that associated with the formation of
condensates of one or more gaugino bilinears in the confining phase of hidden
Yang-Mills sector \cite {nilles}. A non-zero
expectation value for a gaugino bilinear does not in itself necessarily lead to
SUSY-breaking
\cite{veneziano}: the mechanism hinges on the presence of  an additional hidden
sector field ({\em e.g.} the dilaton) in the kinetic function of the gauge
field. In this case, the gaugino bilinear can indeed generate a non-zero
$F$-term for the other field, and thus break SUSY. Of course, many problems
remain, such as  the absence of physically well-motivated models, and the
mechanism for the cancellation of the cosmological constant. Nevertheless, this
picture does provide at least a heuristic focus for discussion.

The usual approach, in such discussions of SUSY-breaking, is to begin with the
condensate at (or near) its stable-vacuum value, and to integrate it out, as a
constraint field, in favor of the light dynamical variables (dilaton, moduli)
whose further evolution one may wish to study. In this work, I will focus on
the
process of formation of the condensate itself, using an effective field theory
in the framework of the standard cosmology. It will be seen that the anomaly
constraints on the effective theory, delineated very early in the discussion of
this topic by Veneziano and Yankielowicz \cite{veneziano}, will play the
determining role in the problems encountered. The global SUSY discussion in
Ref.~\cite{veneziano} need only be amended by more recent developments
concerning SUSY-breaking in order to arrive at the major result of interest:
in the weak coupling domain of the dilaton, the formation of the condensate is
exponentially suppressed by a potential barrier whose origin are the anomalies
of the supersymmetric Yang-Mills theory corresponding to scale and $U_X(1)$
(axial) transformations. (The latter is  an $R$-type  symmetry.) Completion of
the phase transition suffers from the same lack of a `graceful exit' which
proved so problematic for  the original inflationary paradigm.
\bigskip

\section{The Effective Lagrangian}

The dynamical framework of the discussion which follows is that of a
supersymmetric Yang-Mills theory in the confining phase, supplemented
by the presence of the `stringy' objects, the dilaton $S$ and the moduli
$\mod.$ (The symbol $T$ will be reserved for temperature.) Following
Refs.~\cite{veneziano} and
\cite{ferrara}, I will consider an effective Lagrangian based on the
superpotential (for $SU(N)$)
\be
W=\tfrac{1}{32\pi^2}\ U\ \log \left(U^{N}\ e^{8\pi^2S}\right)\ \times
\mbox{modular function}\ \ ,
\label{eq:w}
\ee
where $U$ is the chiral superfield  whose first component (also called
$U$) is the gaugino bilinear. The effective Lagrangian will depend on the
Kahler
form
$K(\Us,U)$ for the $U$-field:

\beq {\cal L}_{eff}&=&{\cal L}_{kin} - V\nonumber\\
{\cal L}_{kin}&=&
K_{\Us U}\ \partial^{\mu}\Us\partial_{\mu}U +
\mbox{ kinetic terms for\ } S, \mod,\ldots\nonumber\\
V&=&\mbox{(function of $S, \mod$ )}\ \cdot
K^{-1}_{\Us U}\
\log(\Us/{\mu\pri}^3)^N\ \log(U/{\mu\pri}^3)^N\nonumber \\
&&\hspace*{1cm}-\mbox  {terms of }
O(\uu)/M^2\  \ .
\label{eq:leffu}
\eeq In Eq.~(\ref{eq:leffu}),
\bi
\item $K_{\Us U}=\partial^2K/\partial\Us\partial U$
\item $M=M_{Planck}$
\item $\mu\pri\simeq Me^{-8\pi^2S/3N}$
\item  the second term in $V$ constitutes the sum of the additional  \s\ and
$\mod$ $F$-terms, as well as the $-3|W|^2/M^2$ term of the supergravity
potential.
\ei

 The effect of the second term in $V$ is crucial: it will (one hopes!)
displace the
potential  minimum at $U={\mu\pri}^3$ below that at
$U=0,$ and, with a zero cosmological constant, the needed cancellation will
require a non-zero $F$-term, consequently breaking supersymmetry. In this paper
no proposals are made for the machinery to accomplish these purposes;
a viable model is simply asumed to exist, and the consequences are
examined.

In order to proceed, the properties of the Kahler potential need to be
specified. In this paper, the initial (and primary) focus will be on the scale
invariant Kahler proposed in \cite{veneziano}
\be
K(\Us,U)= a (\uu)^{1/3}\ \ ,
\label{eq:kahlersc}
\ee
where $a$ is a positive constant.
This will provide an intuitively reasonable interpolation  between the high and
low temperature phases. It will, however, present ambiguities with respect to
Witten's index theorem \cite{witten}. A `softened' form of the Kahler,
\be
K(\Us,U)\sim \uu/\mu^4\ \ ,
%K(\Us,U)\sim \mu^2 (\uu/\mu^6)^p\ , \quad p>\tfrac{1}{3}\ ,\ \ ,
\label{eq:kahlersoft}
\ee
will also be discussed in Section 5 as an illustrative alternative:
it will be seen to   generate a dynamics whose high
temperature phase is difficult to interpret.
%will also be discussed. For $p<\half,$ the dynamics will be similar to the
%%case
%$p=\frac{1}{3};$ for $p\geq\half,$ the
%high temperature phase   of the resulting
%theory will be  difficult to interpret.

With the Kahler form (\ref{eq:kahlersc}) it is convenient
to rescale \cite{veneziano,ferrara} to a field
$Y$ with canonical dimensions via $U=(Y/\sqrt{a})^3,$ to obtain a
working version of ${\cal L}_{eff}:$
\beq {\cal L}_{eff}&=& \partial^{\mu}\Ys\partial_{\mu}Y +\mbox{kinetic terms
for\ } S,\mod\;\; - V(Y,S,\mod)\nonumber\\[2mm]
V(Y,S,\mod)&=&f(S,\mod)\
\cdot\
(\YY)^2\ \log(Y/\mu)^{3N}\ \log({\Ys}/\mu)^{3N}\nonumber\\[2mm]
&&\hspace*{2cm} - \mbox{terms of }O((\YY)^3/M^2)\ ,
\label{eq:leffy}
\eeq where
\be
\mu=\sqrt{a}\mu\pri\sim M e^{-8\pi^2S/3N}\ \ .
\label{eq:mu}
\ee

In the absence of other fields, the minima at $Ye^{2\pi ik/3N}=\mu$ do not
break
super\-symmetry\cite{veneziano}. The remaining terms will  displace
this minimum downward by an amount
\be
\eps\sim \mu^6/M^2\sim \ms^2M^2
\label{eq:eps}
\ee representing the difference between vacuum energies of the broken and
unbroken  phases. Once this minimum has been attained, further evolution in the
$S$ and $\mod$ variables will drive $V$ to its global minimum.

The potential is not invariant under the $U(1)$ rotation of $Y,$ reflecting the
$U_X(1)$  anomaly; however, it does display the $Z(N)$ symmetry for $|Y|=\mu.$
For real
$\mu,$ the minimum of the potential lies very near $|Y|=\mu,\ $
(displaced slightly by the last term in
\eqn{leffy}), and in what follows I will consider the evolution of
$Y$ along the real axis. A sketch of the potential along the real $Y$-axis, for
fixed \s\ and $\mod,$ and  with the
SUSY-breaking gap $\eps$ greatly exaggerated,
is shown as the solid line in Fig. 1. Central to the development of the paper
is a discussion, which follows, of the zero of the potential at $Y=0.$

\section{The Configuration $Y=0.$}

In the global supersymmetric limit $(M=0)$ the potential shows a
field configuration at $Y=0$ which is degenerate with the $N$ vacua at
$|Y|=\mu.$ What is the meaning of this configuration? It is clearly
tied to the choice of Kahler, since the `softened' choice
(\ref{eq:kahlersoft}) will cause $V(U=0)\neq 0,$ with the same
superpotential. The Witten index for
$SU(N)$ is $N$ \cite{witten}, which is the number of vacua at
$|Y|=\mu.$
It may be, then, that the effective theory (with the Kahler
(\ref{eq:kahlersc}))  should be modified in order to support a
breaking of the degeneracy between
$Y=0$ and
$Y=\mu.$ However, requiring that the discussion of the phase transition
make use of conventional field-theoretic methods places several constraints
on the dynamical framework:
\bi
\item the model should contain attainable field configurations with both zero
and non-zero condensate;
\item the model should allow a study of the phase transition
using
canonical finite temperature corrections;
\item the finite temperature corrections should localize the field
configuration to zero condensate at high temperature.
\ei
The finite-temperature discussion in the next section will show these
requirements to be met by the Lagrangian (\ref{eq:leffy}). It will also be
pointed out, in a parallel fashion, why the use of the softened Kahler
(\ref{eq:kahlersoft}) fails to meet the criteria above. Since a finite
temperature environment breaks supersymmetry, one should perhaps not worry
about the seemingly extra zero-temperature vacuum at $Y=0$, as long as $T\neq
0.$ This will be the philosophy followed in this work. The transition to
$T\simeq 0$ will be carried out as a continuation from $T\neq 0.$

\section{Evolution of the Condensate}

A salient feature of the potential $V$ can be immediately
noted: {\em if} the gaugino bilinear is near zero  at early times, then the
stable vacuum at $|Y|=\mu$ can  be attained only
\bi
\item[$(a)$] by thermal excitation over the barrier;
\item[$(b)$] via quantum tunneling through the barrier;
\item[$(c)$] by a classical roll if the condition $Y=\mu(S,\mod)$ is obeyed at
{\em all} points along the roll. (This eliminates the barrier: see
\eqn{leffy}.)
\ei

It is this transition from zero to non-zero condensate, in a cosmological
context, which is the subject of this study. The formation of critical
bubbles of true vacuum via either thermal excitation or quantum tunneling
(cases $(a)$ and $(b)$ above) is controlled by actions calculated from either
$O(3)$- or
$O(4)$-invariant solutions to the classical field equations; thus, the
simultaneous behavior in the complete field space of $Y, S$ and the moduli is
technically required. This is difficult to implement; instead, I will calculate
the transition in the
$Y$-variable for some fixed average value of $S$ (ignoring the moduli),
and extract
{\em a posteriori} the constraint on this average value for the validity of the
result. The single exception to this, case
$(c)$, will receive separate comment in Section 8.
Finally, because the only two mass scales are $\mu$ and
$M,$ I will not retain  numerical factors of $O(1)$ in what follows, since
they will be irrelevant in exploring the parametric dependence of the results
on
$\mu$ and
$M$.

\section{Critical Bubble Formation: $T\not = 0.$}

In order to discuss the finite temperature situation, I now give a brief
sketch of the thermal environment which will be assumed for
the situation at hand. Standard finite temperature  field theory concepts will
be used since the strength of the interactions is sufficient
to maintain thermal equilibrium.

At temperatures
$T\gg\mu\ $, the perturbative description of the theory is
appropriate, so that $V$ in
\eqn{leffu} is not meaningful. When $T\sim \mu,$ the effective theory becomes
applicable, with the zero-temperature potential given in
(\ref{eq:leffy}) supplemented by the one-loop contribution
\cite{jackiw,weinberg} to the free energy
\beq
\Delta V_T(Y,T)&\sim& T^4\int^{\infty}_0 dx\ x^2 \
\ln(1-e^{-\sqrt{x^2+m^2(Y)/T^2}}\nonumber\\
&+&\mbox{similar term for fermions} + C\ \ .
\label{eq:vty}
\eeq
Here $m^2(Y)=\partial^2V/\partial Y^2,$ and the real part of $\dvt$ will be
used
when $m^2<0.$ The constant $C$ will eventually
be fixed so that $\dvt(Y=0)=0;$ thus, $C\sim +T^4.$

Details of $\dvt$  will not be required in the present consideration. It will
suffice to approximate it (its real part, at least) by a function
which runs from
\be
\Delta V_T(Y,T)\sim -T^4+T^2Y^2 + C \sim T^2Y^2
\label{eq:lowy}
\ee
near
$Y=0$ (the free energy of massless quanta in the false vacuum + large $T$
correction) to
\be
\Delta V_T(Y,T)\sim 0 + C \sim T^4
\label{eq:highy}
\ee
 near
$Y=\mu$ (only massive glueballs or `hidden hadrons' in the
true vacuum). The free
energy of other (non-hidden) fields which interact only gravitationally with
the hidden gauge fields will play no role in this discussion. The dashed curve
in Fig.~1 represents $V(Y,T)=V(Y)+\Delta V_T(Y,T),$ the sum of the zero
temperature potential and the correction. With these minimal
properties, two observations are important:
\bi
\item[(1)] For $T\sim\mu,$ the confining properties of the second term
on the r.h.s. of \eqn{lowy} are expected to force the condensate to
begin its evolution near $Y=0;$
\item[(2)]
no phase transition can occur until
$T^4\lsim \eps\ .$ This defines a critical temperature
\be
T_c\sim \eps^{1/4}\sim \mu^{3/2}/M^{1/2}\sim 10^{12}\ \gev\ \ .
\ee
The dashed curve in Fig.~1 is appropriate to $T>T_c.$
\ei
Thus, it is seen that the effective theory with the Kahler (\ref{eq:kahlersc})
satisfies the constraints itemized in Section 3. If one  instead
attempts a finite temperature modification of the effective theory with the
softened Kahler (\ref{eq:kahlersoft}) (this time with $U$ as the field
variable), it is easy to check that $(a)$ the potential is singular at $U=0;$
$(b)$ the finite temperature correction is negligible everywhere in field space
(the effective mass $m^2(U)=\partial^2V/\partial U^2>\mu>T$ everywhere). Thus,
there is no mechanism to support as initial condition $U=0$ in this theory, and
it fails to meet the criteria for interpolation between $U=0$ and $U=\mu^3$
delineated in Section 3.

The calculation of the nucleation rate now proceeds along
standard lines
\cite{linde}:
one evaluates the stationary O(3)-invariant  (Euclidean)
action ({\em i.e.,} the bubble energy) for the appropriate boundary conditions,
and divides it by
$T$, in order to obtain the Boltzmann exponent. Because of the large barrier
and the small energy gap between false and true vacua,
thin-walled bubbles are optimal \cite{coleman}, so that, for $T<T_c,$ the
three-dimensional action (the  difference between true and false vacuum
energies in a bubble of radius $R$) becomes, as a function of $R,$

\be
E(T)\approx -R^3\eps + R^2 S_1(T)\ \ ,
\label{eq:eb}
\ee
where
$\eps$ was given in \eqn{eps}, and
$S_1(T)$ is the one-dimensional action integrated through the bubble wall
\cite{coleman},
\be
S_1(T)=\int^{Y^-}_{Y_+}\ dY\ [2(V(Y,T)]^{1/2}\ \ ,
\label{eq:s1t}
\ee
with $V(Y,T)=V(Y)+\Delta V_T.$
(In practice, we may take $Y_-=0,\ Y_+=\mu.$)
Varying  $E(T)$ with respect to
$R$ gives for the critical bubble
\beq R(T)&\sim& S_1(T)/\eps\ \ ,\nonumber\\ E(T)&\sim& S_1(T)^3/
\eps^2\ \ .
\label{eq:ret}
\eeq
At the low temperature $T_c$ there is negligible error in omitting
the contribution of $\dvt$ in calculating the surface term $S_1(T).$
With the use of \eqn{leffy}, one obtains
\be
S_1(T<T_c)\simeq S_1(0)\sim \mu^3 \ \ .
\label{eq:stc}
\ee
{}From Eqs.~(\ref{eq:ret}),(\ref{eq:stc}),
the critical bubble energy  for $T<T_c$ is
\be
E(T<T_c)\sim E(0)\sim \mu^9/\eps^2\sim \mu\ (M/\mu)^4\ \ .
\label{eq:et}
\ee
The bubble radius at nucleation is
\be
R(T<T_c)\sim R(0)\sim \mu^3/\eps\sim M^2/\mu^3\ \ ,
\label{eq:rt}
\ee
which is just the de Sitter horizon $H^{-1}$ corresponding to the vacuum energy
$\eps.$ Thus, general relativistic corrections are expected to be of $O(1)$ at
most. (Further discussion will be given in the next section.) At this point, I
also note that the thin-wall approximation is justified: the thickness of
the bubble wall $\Delta R\sim \mu^{-1}$ \cite{coleman} is much smaller than
the the bubble radius $R\sim M^2/\mu^3.$

The Boltzmann exponent controlling thermal bubble nucleation for $T<T_c$ is
\be
B(T)\simeq
E(0)/T\ge (M^4/\mu^3)/T_c\sim (M/\mu)^{4.5}\ \ .
\label{eq:bt}
\ee
which gives  an average nucleation rate/unit volume of
\be
\Gamma/V\le \mu^4\ e^{-B(T_c)}\ \ .
\label{eq:gam}
\ee

It is clear from \eqn{bt} that the Boltzmann exponent is huge unless $\mu\gsim
0.1M.$ From \eqn{mu} this proximity of $\mu$ and $M$ is possible only if $S$ is
in a domain corresponding to strong gauge coupling at
the time of the nucleation.
A compelling argument that this is not the case has been made by Brustein and
Steinhardt
\cite{brustein}: if the dilaton  were indeed  in the
strong (gauge) coupling regime $S\lsim 1$ at the time of
gaugino condensation, its  large potential energy in
this configuration would carry it quickly to
$S\rightarrow \infty,$ beyond its stable vacuum value. If it is assumed that
this is not the case, then the large  result for
$B$ indicates {\em ipso facto} that the
$T\not = 0$ calculation represents an upper bound on the transition rate: for
$T<T_c$  the  energy density of the universe is dominated by the vacuum energy
$\eps,$ and the temperature will cool to zero in a few Hubble times of the de
Sitter expansion, much smaller than the average time
$t_N$ to nucleate a bubble:
\be t_N\sim (\Gamma/V\cdot H^{-3})^{-1}\sim H^{-1}\ (H/\mu)^4\
e^{B(T)}\sim H^{-1}(\mu/M)^8\ e^{B(T)}\gg H^{-1}\ \ .
\label{eq:tn}
\ee Thus, cooling to $T\simeq 0$ takes place well before any thermal nucleation
out of the false vacuum. In order to circumvent the ambiguity (raised in
Section 3) of reconciling the configuration $Y=0$
with the zero-temperature index
theorem, I will keep the temperature small but non-zero $(T\ll T_c)$. This will
allow a calculation of  the quantum tunneling rate in a zero-temperature
formalism.

\section{Quantum Tunneling: $T=0.$}

The calculational procedure here is well-known \cite{coleman}: the decay of the
false vacuum proceeds via  the O(4) invariant Euclidean instanton
solution connecting
$Y\simeq 0$ and the appropriate point in $Y$ beyond the barrier.
 For the same reason given in the O(3) discussion of the
last section,  the thin-wall
approximation
\cite{coleman} for bubble formation will be appropriate.

The zero temperature decay rate of the false vacuum is given by
\be
\Gamma/V\sim \mu^4\ e^{-B_0}\ \ ,
\label{eq:gam0}
\ee
where $B_0$ is the O(4) bounce. From \cite{coleman}, this is (in the thin-wall
approximation)
\be
B_0\sim S_1(0)^4/\eps^3\sim (M/\mu)^6\ (\sim 10^{33}\ \mbox{for $\mu\simeq$ its
true vacuum value)}\ \ ,
\label{eq:b0}
\ee
where  the zero temperature surface energy $S_1(0)$ was estimated in
\eqn{stc} \cite{s1}. The bubble size is the same as estimated in
\eqn{rt},
\be
R\sim M^2/\mu^3\sim H^{-1}\ \ ,
\label{eq:r0}
\ee
the horizon size. Again, general relativistic corrections can be expected to be
of $O(1)$\cite{genrel}.
\section{Implications of Eqs.~(21) and (22)}

Barring strong coupling (see discussion following \eqn{gam}),  the problem
posed by the results (\ref{eq:gam0}) and (\ref{eq:b0}) is  clear: associated
with excessive supercooling, there result clustering inhomogeneities of true
vacuum \cite{guthwein}  which are incompatible with the observed smoothness of
the visible universe.  In order to avoid  these inhomogeneities, it is
necessary
that the probability
$\eta$ to nucleate a true vacuum bubble in a Hubble  space-time volume not be
$\ll 1.$
\cite{guthwein}. In the present case, unfortunately,
\be
\eta=\Gamma/V\cdot H^{-4}\sim (M/\mu)^8 \ e^{-(M/\mu)^6}\sim 10^{-10^{30}}\ \ .
\label{eq:pbubble}
\ee Thus, if none of the hypotheses which led to \eqn{b0} is modified, there
appears to be a cosmological problem associated with  hidden gaugino
condensation as  the progenitor to SUSY-breaking.  In the concluding section, I
will  review  the input which led to the result, and comment on some possible
modifications which need to be explored.

\section{Classical Roll}

This possibility was mentioned as a technical option in the introductory
section. A glance at the potential
$V(Y)$ in
\eqn{leffy} shows that the barrier would be absent for a path in field space
$Y=\mu(S).$ (I will omit the moduli from the discussion for purposes of
simplicity.) Such a path obviously exists, but  its probability will be
exponentially suppressed unless it is a solution to the classical equations.
This requirement imposes very special conditions on the dynamical variables
(such as  the Kahler potential for the dilaton); at present these seem
somewhat {\em ad hoc}, but perhaps this option merits further study.

\section{Remarks and Summary}

\ni{1}The principal source of difficulty for formation of the condensate is the
presence of $(\log Y)^2$ potential barrier in
\eqn{leffy}, which is a direct result of implementing the proper scale and
axial
anomalies in the effective theory. However, explicit implementation is not
required in working  with the elementary fields \cite{dds}. Of course,  we have
at present no method of calculating the evolution of the condensate  through
the
phase transition using the elementary theory. In QCD with light quarks, there
has been some evidence that an  effective theory ({\em i.e.,} the sigma model
\cite{hg,pisarski}) can successfully  model  the chiral phase transition
subsequently observed in lattice QCD
\cite{gupta}.

\ni{2}The results of this work carry over to the case of more than one
condensate, since the condensates for product groups evolve
independently.

\ni{3}  Matter fields have thus far not been included in the analysis, and I
have not seen any simple way of characterizing their possible influence on the
result.  Their inclusion \cite{tvy,affleck}
complicates the picture enormously, especially if they are
massless
\cite{affleck,amati}. In the case of bilinear
condensates $\pij=Q^i\bar Q_j$ of
fields in
$N+\bar N$ representations of $SU(N)$, there is non-trivial kinetic mixing of
the $\pij$ with the gaugino bilinear $U$ \cite{tvy}\cite{remark}. It is
unclear whether this mixing  could eventually destabilize
the region of $U\simeq 0$ as a possible starting point for the evolution of the
gaugino condensate. If it does not, then there would seem to be no immediate
reason to modify the conclusions of this paper.

\ni{4}To summarize, the principal result of this work is that the effective
theory describing gaugino condensation also creates a barrier to a successful
completion of the phase transition from zero to non-zero condensate. Only if
the
gauge theory is strongly coupled at the time of the transition (so that
$\mu(S)\sim M)$ can this difficulty be circumvented. But this would come at the
price of unacceptable behavior for the dilaton after the phase transition. If
gaugino condensation should become unfavored as a candidate to provide a
superpotential for moduli, then perhaps matter condensates, in combination
with stringy non-perturbative modifications
manifested directly in the Kahler potentials of the moduli
\cite{banksdine,banks}, are responsible for
SUSY-breaking.

I would like to thank Tom Taylor, Gabriele Veneziano, and Misha Voloshin
for useful discussions during the
course of this work, and for focussing attention on the questions
surrounding the vacuum at $Y=0$ in the effective Lagrangian.
This work was supported in part by Grant No. PHY-9411546 from
the National Science Foundation.

\section*{Figure Caption}
\noindent \parbox[t]{0.5in}{Fig.1}\parbox[t]{6.0in}{ Solid Curve: the zero
temperature potential;
\\ Dashed curve: the finite temperature potential for
$T>T_c,$  normalized to 0 at
$Y=0.$}

\begin{thebibliography}{99}
\bibitem{nilles}S.~Ferrara, L.~Girardello, and H.~P.~Nilles,
\plett{125}457 (1983);  J.~-~P.~Derendinger, L.~E.~Ib\`a\~nez, and
H.~P.~Nilles,
\plett{155} (1985) 65;  M.~Dine, R.~Rohm, N.~Seiberg, and E.~Witten,
\plett{156} (1985) 55;  C.~Kounnas and M.~Porrati, \plett{191} (1987) 91;
S.~Ferrara, N.~Magnoli, T.~R.~Taylor, and G.~Veneziano,
\plett{245} (1990) 243; A.~Font, L.~Ib\`a\~nez, D.~Lust, and F.~Quevedo,
\plett{245} (1990) 401; H.~P.~Nilles and M.~Olechowski, \plett{248} (1990) 268;
P.~Bin\'etruy and M.~K.~Gaillard, \plett{253} (1991) 119;
\np{358}(1991)194; B.~de~Carlos,  J.~A.~Casas, and C.~Mu\~noz, \np{399}623
(1993).
\bibitem{veneziano}G.~Veneziano and S.~Yankielowicz,
\plett{113}231 (1982).
\bibitem{ferrara}S.~Ferrara, N.~Magnoli, T.~R.~Taylor, and G.~Veneziano,
Ref.~\cite{nilles}
\bibitem{witten}E.~Witten, \np{202} (1982) 253.
\bibitem{jackiw}L.~Dolan and R.~Jackiw, \prd{9} (1974) 3320.
\bibitem{weinberg}S.~Weinberg, \prd{9} (1974) 3357.
\bibitem{linde}A.~Linde, \plett{70} (1977) 306; {\em ibid} {\bf 92B} (1980)
119.
\bibitem{coleman}S.~Coleman, \prd{15} (1977) 2929; C.~Callan and S.~Coleman,
{\em ibid.} {\bf D16} (1977) 1762.
\bibitem{brustein}R.~Brustein and P.~J.~Steinhardt, \plett {302} (1993) 196.
\bibitem{s1}The one-dimensional action (and the bounce) are the same in this
case as would be expected for tunneling between local minima in a generic
non-perturbative dilaton potential \cite{banks}: in that case, the potential
barrier is lower
$(\sim
\mu^6/M^2$ {\em vs.}
$\mu^4$ in the present case); however, the width is larger $(\sim M$ {\em vs.}
$\mu$ in the present case).
\bibitem{genrel}The effect on the bounce of including general relativity in the
analysis was shown by Coleman and De Luccia
\cite{colemandel} to constitute a multiplicative factor of
$[1+(3S_1^2)/(4\eps M^2)]^2\sim O(1)$ in the present case. This is so because
the  bubble radius $R\sim H^{-1}$, as calculated without metric corrections;
the
Coleman-DeLuccia construction (which ensures that $R\le H^{-1})$ then provides
a
modification which is parametrically of $O(1).$
\bibitem{colemandel}S.~Coleman and F.~ De Luccia, \prd{21}(1980) 3305.
\bibitem{guthwein}A.~H.~Guth and E.~J.~Weinberg, \prd{23} (1981) 876; \np{212}
(1983) 321; S.~W.~Hawking, I.~G.~Moss, and J.~M.~Stewart, \prd{26} (1982) 2681.
\bibitem{dds}A.~C.~Davis, M.~Dine and N.~Seiberg, \plett{125} (1983) 487.
\bibitem{hg}H.~Goldberg,  \plett {131} (1983) 133.
\bibitem{pisarski}R.~D.~Pisarski and F.~Wilczek, \prd {29} (1984) 338.
\bibitem{gupta}R.~Gupta, G.~Guralnik, G.~Kilcup, A.~Patel,  and S.~R.~Sharpe,
\prl {57} (1986) 2621; J.~B.~Kogut, H.~W.~Wyld, F.~Karsch, and D.~K.~Sinclair,
\plett {188} (1987) 353; M.~Fukugita, S.~Ohta, Y.~Oyanagi, and A.~Ukawa, \prl
{58} (1987) 2515.
\bibitem{tvy}T.~R.~Taylor, G.~Veneziano, and S.~Yankielowicz,
\np{218}(1983) 493.
\bibitem{affleck}I.~Affleck, M.~Dine, and N.~Seiberg,
\np{241}(1984) 493;
\np{256}(1985) 557.
\bibitem{amati}D. Amati, K. Konishi, Y. Meurice, G.C. Rossi and G. Veneziano,
{\em Physics Reports} {\bf 162} (1988) 169.
\bibitem{remark}Recall that since the evolution of the condensate toward its
non-zero value is being studied, one cannot follow the normal procedure of
integrating out the heavy condensate field to leave an effective theory of
light
composite matter fields.
\bibitem{banksdine}T.~Banks and M.~Dine, \prd{50}(1994) 7454.
\bibitem{banks} T.~Banks, M.~Berkooz, S.~H.~Shenker, G.~Moore, and
P.~J.~Steinhardt, hep-th/9503114, RU-94-93.

\eb\ed
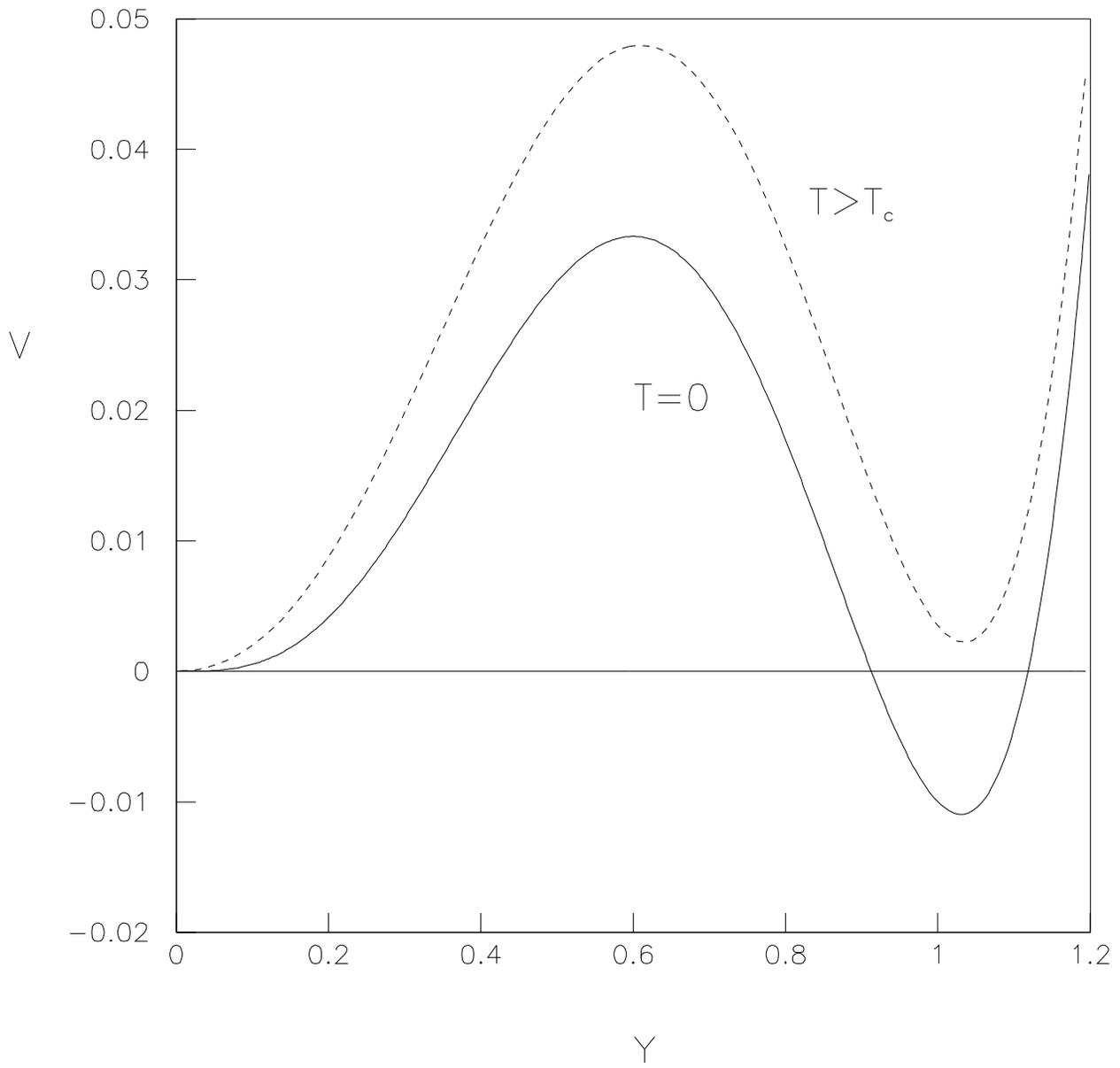
